\documentclass{article}
\usepackage{latexsym, amsmath}

\everydisplay{\rm }
\everymath{\rm}

\begin{document}
\begin{center}
\LARGE Planar spin network coherent states  \\ [.125in] II. Small
corrections
\\ [.25in] \large Donald E. Neville \footnote{\large Electronic
address: dneville@temple.edu}
\\Department of Physics
\\Temple University
\\Philadelphia 19122, Pa. \\ [.25in]
April 29, 2013 \\ [.25in]
\end{center}

\newcommand{\E}[2]{\mbox{$\tilde{{\rm E}} ^{#1}_{#2}$}}
\newcommand{\A}[2]{\mbox{${\rm A}^{#1}_{#2}$}}
\newcommand{\Np}{\mbox{${\rm N}'$}}
\newcommand{\Etwo}{\mbox{$^{(2)}\!\tilde{\rm E} $}\ }
\newcommand{\Etld }{\mbox{$\tilde{\rm E}  $}\ }
\newcommand{\Vtwosq}{\mbox{$(^{(2)}\!{\rm V})^2 $}\,}
\newcommand{\Vtwo}{\mbox{$^{(2)}\!{\rm V} $}\,}
\newcommand{\Vthree}{\mbox{$^{(3)}\!{\rm V} $}\,}

\newcommand{\bea}{\begin{eqnarray}}
\newcommand{\eea}{\end{eqnarray}}
\newcommand{\be}{\begin{equation}}
\newcommand{\ee}{\end{equation}}
\newcommand{\nn}{\nonumber \\}
\newcommand{\rta}{\mbox{$\rightarrow$}}
\newcommand{\rla}{\mbox{$\leftrightarrow$}}
\newcommand{\eq}[1]{eq.~(\ref{#1})}
\newcommand{\Eq}[1]{Eq.~(\ref{#1})}
\newcommand{\eqs}[2]{eqs.~(\ref{#1}) and (\ref{#2})}
\newcommand{\Eqs}[2]{Eqs.~(\ref{#1}) and (\ref{#2})}
\newcommand{\bra}[1]{\langle #1 \mid}
\newcommand{\ket}[1]{\mid #1 \rangle}
\newcommand{\braket}[2]{\langle #1 \mid #2 \rangle}
\large
\begin{center}
{\bf Abstract}
\end{center}
    This paper is the second of two which construct
coherent states for spin networks with
planar symmetry.
Paper 1 constructs a set of
coherent states peaked at specific values of
holonomy and triad.  These operators acting
on the coherent state give back the coherent
state plus small correction (SC) states.  The present
paper proves that the SC states form a complete subset
of the overcomplete set of coherent states.  The
subset is used to construct a perturbation expansion
of the inverse of the volume operator.  Appendices
calculate the standard deviations of the
angles occurring in the holonomies,  demonstrate
that standard deviations are given by matrix
elements of the SC states, and estimate the rate
of spreading of a coherent state wave packet.
\\[.125in]
PACS categories: 04.60, 04.30

\clearpage
\section{Introduction}\label{SecIntroduction}

    This paper is the second of two which construct
coherent states for spin networks with
planar symmetry.  The first paper of this series
(which will be referred to as paper 1)
constructed a set of
coherent states for planar gravity waves \cite{PCSI}.
When holonomy and triad operators act on these states,
one gets the original state, plus small
correction (SC) states.  Those states were studied
in the appendices to paper 1.

    The coherent states form an overcomplete set, and as such are
not well suited as a basis for perturbation theory.  Section \ref{Complete}
constructs a complete subset of the overcomplete set. Sections \ref{V2Squared},
\ref{V2}, and \ref{V2Inverse} use the
subset to construct series expansions for, respectively,
the square of the volume operator, the volume operator,
and the inverse of the volume operator.

    The following is a summary of notation from
paper 1.  The coherent states are sums of O(3) spherical
harmonics.
\begin{multline}
    \ket{ u,\vec{p}} = N \sum_{L,M}((2L+1)/4\pi) \\
                   \cdot \exp[-t L(L+1)/2 \,] D^{(L)}(h)_{0M}D^{(L)}(g\dag)_{M0}.
                                \label{defcoh}
\end{multline}
The rotation matrix D(h) is a spherical harmonic
$Y_{LM}$, apart from normalization.
\begin{equation}
    \sqrt{4\pi/(2L+1)}Y_{LM}(\theta /2,\phi -\pi/2) =
            D^{(L)}(h)_{0M}(-\phi +\pi /2, \theta /2,\phi -\pi/2).
                \label{YeqD}
\end{equation}
The matrix g, \eq{defcoh}, is in SL(2,C); every such matrix can
be decomposed into a product
of a Hermitean matrix H times a unitary matrix u.
\bea
    g^{(L)} &=& \exp [\,\vec{S}\cdot \vec{p} \,] \,u^{(L)} \nn
        & := & H^{(L)}\, u^{(L)}
        \label{geqHu}
\eea
All three
matrices (h, H, and u) have their axis of rotation in the XY plane.
\begin{eqnarray}
    h^{(L)} & = & \exp[ \,i \, \hat{m} \cdot \vec{S} \,\theta/2 \,] ; \label{defh}\\
    \hat{m} &=& (\cos \phi, \,\sin \phi, \,0).
        \label{defm}
\end{eqnarray}
\bea
      u^{(L)} &:=& \exp[\,i \, \hat{n} \cdot \vec{S}\, \alpha /2 \,];\nn
    \hat{n} &=& (\cos \beta, \,\sin \beta, \,0).
           \label{defu}
\eea
\bea
    H^{(L)}  &=& \exp [\,\vec{S}\cdot \vec{p} \,]; \nn
       \vec{p}& =& p \,[ \,\cos(\beta +\mu), \,\sin(\beta +\mu), \,0 \,].
        \label{defmu}
\eea
The matrix u determines the peak value of the
holonomy h, while $\vec{p}$ determines the
peak angular momentum.

    The Hamiltonian is expected to contain both
densitized triads \Etld and holonomies
$h^{(1/2)}$ in the fundamental representation of
SU(2).  However, the action of the \Etld on the O(3)
harmonics of \eqs{defcoh}{YeqD} is
simpler than the action on products of the $h^{(1/2)}$.
Therefore it is desirable to work entirely with
O(3) harmonics, including replacing the
SU(2) representations in the Hamiltonian by O(3)
harmonics.  Each $h^{(1/2)}_{mn}$ in the
Hamiltonian may be replaced by an O(3)
spherical harmonic, by using the formulas
\bea
    \sqrt{4\pi/(3)}Y_{1M} &=& D^{(1)}(h)_{0M} := \hat{\mathrm{D}}_M(h); \nn
    \hat{D}(h)_{\pm} &=& h^{(1/2)}_{\mp\,\pm}/\sqrt{2};\nn
    \hat{D}(h)_0 & = & h^{(1/2)}_{++} = h^{(1/2)}_{--}.
    \label{Veqh}
\eea

    The direction of \Etld is given by
$\hat{p}$ rotated out of the XY plane by
the rotation u.
\begin{eqnarray}
    (\gamma\kappa/2)^{-1}\E{x}{A} \ket{u,\vec{p}} &=&
            <L + 1/2> (\hat{n}_A \cos \mu -(\hat{n}\times \hat{D})_A \sin \mu)\ket{u,\vec{p}}
                + SC \nn
                &=& <L +1/2> \hat{p}_B D^{(1)}(u)_{BA} \ket{u,\vec{p}}+ SC \nn
                &:=& <L +1/2> \hat{L}_A \ket{u,\vec{p}}+ SC
                \label{Eeigenval}
\end{eqnarray}
$\hat{D}$ is now $\hat{D}(u)$, i.~e. $\hat{D}(h)$ evaluated at
the peak value of the holonomy h.

    The $H^{(L)}$ factor, \eq{defmu}, diverges as $\exp(p\,L)$;
together with the explicit exponential in \eq{defcoh},
this gives a peak value of L.
\[
    <\sqrt{L(L+1)}> \,\cong \,<L +1/2> = p/t.
\]

\section{Small Parameters}

        The properties of the coherent states are
derived in paper 1 using two approximations.  Corresponding
to the  two approximations, there are two  small
parameters, $e^{-p}$ and $1/\sqrt{<L>}$.
Higher powers of $e^{-p}$ are neglected
in the appendix on D(H), paper 1; the result is a
manageable expression for D(H).  The SC terms are down
by powers of the second small parameter, $1/\sqrt{<L>}$.  We wish to include
the SC terms, but continue to drop
terms which  are down by $e^{-p}$. This requires
\begin{equation}
    e^{-p} << 1/\sqrt{2 \,p \,<L>}.
\label{InequalityI}
\end{equation}
I include an additional factor of $\sqrt{2p}$;
some SC terms are down by that additional
factor.  \Eq{InequalityI} may be rewritten
\begin{equation}
    e^{2p}/2p := L_0 >> L.
\label{InequalityII}
\end{equation}
Because of the square root, $L_0$ must be
much larger than L.  I take
\begin{equation}
    max <L> =  L_0/100 =  e^{2p}/200 \,p.
\label{InequalityIII}
\end{equation}
Table \ref{Ta:pvsL} shows values for  p vs. max L.
\begin{table}
\begin{center}
\begin{tabular}{l|c|l}
p   &  max $<L>$ &$\sqrt{ 1/t}$ \\ \hline
6   &136    &4.76 \\
7   &859    &11.1   \\
8   &5,550   &26.4    \\
9   & 36,500 &63.7 \\ \hline
\end{tabular}
\caption{\protect p vs. max $<L>$}
\label{Ta:pvsL}
\end{center}
\end{table}
The last column lists
\[
    \sqrt{ 1/t} = \sqrt{ \mbox{max}\, L/p}.
\]
This quantity is the standard deviation of the Gaussian
distribution of L, with peak value max L;  $\sqrt{1/t}$
should not be large compared to max L.

    The allowable range of L's increases quite rapidly
with p.  Of course there is also a minimum L.  It cannot
get much smaller than 50, or the SC terms ($\sim 1/\sqrt{<L>}$)
will no longer be small.

\section{Constructing a Complete Set}
\label{Complete}

    The standard  perturbation expansion requires
a complete set. Standard theory assumes a
Hamiltonian with an unperturbed part
having known solutions (the complete set),
plus a small perturbing potential.
The set of coherent states
\[
    \{\ket{ u,\vec{p}}, \;\forall \; u,\vec{p} \;\}
\]
is overcomplete,
therefore not suitable for perturbation theory.

    When a triad or holonomy acts on the coherent state
one gets back the coherent state
\begin{equation}
    \Etld \ket{ u,\vec{p}} = \,<\Etld> \, \ket{ u,\vec{p}}
        + SC,
\label{defSC}
\end{equation}
plus small correction (SC) states multiplied by
coefficients which are down by $1/\sqrt{<L>}$.
This section will show that
the states contained in the SC terms, plus
the original state $\ket{u,\vec{p}}$, form a
\emph{complete} subset of the set of coherent states.
The SC term in \eq{defSC} contains the higher order
terms in the perturbation series.  The small expansion parameter
is $1/\sqrt{<L>}$.

    The proof of completeness
uses a fact established in paper 1.  The
action  of \Etld inserts a polynomial linear
in L and M into the summand of the coherent
state,  where M is an internal index (an index summed over
within the state) and (for $\exp(-p)\ll1$)
the rest of the summand equals
two Gaussians in L and M, peaked at $<L> = p/t$
and M = 0 respectively.  Given this
peaking, it is appropriate to shift to the combinations
\[
    <L>, \;L - <L>, \;M.
\]
$<L>$ is a constant which can be taken out of the
summand; this constant gives the leading contribution.

    The terms proportional to L - $<L>$ and M give rise
to the small corrections.  The original Gaussian,
\[
    \exp[-X^2/2\sigma^2],
\]
X = L - $<L>$ or M, is replaced by
\[
    X \exp[-X^2/2\sigma^2].
\]
The SC terms are moments of the Gaussian in the coherent state.
I introduce a notation which emphasizes this.
\be
    \Etld \ket{u,\vec{p}} = \;<\Etld> \;\ket{u,\vec{p}}+
            \sum_X b(1X) \ket{1X},
\label{natlbasis1}
\ee
where $\ket{1X}$ is the original coherent state,
except for an additional factor of X = L - $<L>$ or M.
The "1" in the argument of the ket refers to the first moment.
The "1" in the argument of b refers to the order of magnitude of b.
\begin{equation}
    b(nX) = \mbox{order} <\Etld>/<L>^{n/2}.
\label{OMb}
\end{equation}
These moment states
are of interest because  they provide the complete set.

    \Etld = \E{a}{A} is a vector in Lorentz space.
At the moment I focus on the order of magnitude of each
b and suppress subscripts A, on b and
\Etld. (However, see the next section for the vector
dependence.)

    The process can be continued to generate higher
moments.
\bea
    \Etld \ket{1X} &=& <\Etld > \;\ket{1X} \nn
          & & +  \tilde{b}(1X) \ket{2X} + \tilde{b}(1XX')\ket{1X, 1X'} \nn
          & & +c(1) \ket{u, \vec{p}}.
\label{natlbasis2}
\eea
This result has some surprising features.  $\ket{1X}$
is a coherent state with the same expectation value $<\Etld>$
as the original coherent state $\ket{u,\vec{p}}$, \eq{natlbasis1};
and the "c" term gives back the original coherent state.

    To understand these results, consider X = M for example.
Initial action of \Etld on $\ket{u, \vec{p}} $
inserts an M into the summand.  This splits the
initial Gaussian into two Gaussians.
\[
    \ket{u, \vec{p}} \rta \ket{1M};
\]
\[
    \exp[ \,-M^2/2\sigma^2 \,] \,\rta \,M \exp[\,-M^2/2\sigma^2 \,].
\]
This last expression has a Gaussian peak at M = $\sigma$,
a zero at M = 0, and a Gaussian valley at M = $-\sigma$.
One Gaussian has been turned into (the difference between)
two Gaussians.

    Now let \Etld
act on $\ket{1M}$, \eq{natlbasis2}.  Again, \Etld\emph{}
gives rise to an $<\mathrm{L}>$, L - $<\mathrm{L}>$, and M term in the
summand.   The $<\mathrm{L}>$ can
be taken out of the summand, giving the original
ket  $\ket{1M}$, times the original expectation value $<\Etld>$.
This is the origin of the first term in \eq{natlbasis2}.

    The \Etld also inserts L - $<\mathrm{L}>$  and M into a summand
which already contains an M.
When that initial M was inserted in the summand, it split
one Gaussian into two, with extrema at M = $\pm \sigma$.
The insertion of another factor of M will split each
of those Gaussians  into two more Gaussians.
The initial peak at M = $+\sigma$ becomes a peak and valley
at M $\approx \sigma \pm \sigma = 2 \sigma $(the peak)
and zero (the valley).  The initial valley at M = $-\sigma$
becomes a valley and peak
at M $\approx -\sigma \pm \sigma = 0 $(the valley)
and  $-2 \sigma$ (the peak).  The two peaks at
M $\approx \pm 2 \sigma$ form the second moment
state $\ket{2X}$ in \eq{natlbasis2}; the valley
at M = 0 gives back the original state $\ket{u, \vec{p}}$,
the "c" term in \eq{natlbasis2}.

    When determining the order of magnitude of a matrix element,
one must count the \emph{changes} in moment.  The "c" term
multiplies a ket with zero moments, but is not order zero in
powers of 1/$\sqrt{<L>}$.
It is down by one power of $1/\sqrt{<\mathrm{L}>}$, because on going from
initial to final state the moment
changes by one unit from 1X to 0X.

    There is a similar series of SC states associated with
the holonomy operator.  In this section I consider only the series
generated by the \Etld, because the operators discussed in this
paper depend only on $\tilde{E}$.

    The set of moment states

\be
    [\; \ket{pX, qX'},\:\forall \: \mbox{integers p,q} \, ].
    \label{completeset}
\ee
is complete and can be made orthogonal.
The original coherent state is a Gaussian in M times
a Gaussian in L.
The SC states contain the same Gaussians,
times additional powers of $X-<X>$.  On going from
original coherent state to a SC state, one is therefore
replacing the Gaussian by Gaussian times a sum over
Hermite polynomials in $X - <X>$, i.~e.
a wavefunction for a simple harmonic oscillator.
A ket such as $\ket{pM_x, p'L_x}$, p,p' $\neq $ 0,
contains one sum over Hermite polynomials depending on
$\mathrm{M}_x$ and one sum over polynomials depending on $\mathrm{L}_x$.
The state depends on a matrix H, \eqs{defcoh}{geqHu}, which produces
a Gaussian in $\mathrm{M}_x$ times a Gaussian in $\mathrm{L}_x$.

    When the dot product of two SC states is computed, several factors
drop out, the factors depending on matrices  h and u, \eqs{defcoh}{geqHu}.
The H dependence gives rise to  products of simple harmonic
oscillator wavefunctions in L and M, summed over L and M.  The sums
can be turned into integrals, and the orthonormality
follows.  Completeness follows from the completeness
of the simple harmonic oscillator wavefunctions.

\section{{\boldmath $^{(2)}\!V^2 \:and \: ^{(2)}\!\tilde{E}$}}
\label{V2Squared}

    I begin by calculating the order 1/$\sqrt{<L>}$ terms in
the perturbation series for the action of the volume squared
operator on coherent states.  The matrix of \Etld operators is
block-diagonal because of gauge choices \cite{HusSmo}, with a
1 x 1 subblock containing only \E{z}{Z}.
The volume operator \Vthree
therefore simplifies to
    \bea
        (\Vthree)^2 &=& \mid \E{z}{Z} \, \epsilon_{ZAB} \, \E{x}{A} \, \E{y}{B} \mid \nonumber \\
                &:=& \mid \E{z}{Z} \, \Etwo \mid.\label{V3}
    \eea

    The holonomies along the z direction are eigenstates of  \E{z}{Z};
computing square roots and inverses is relatively simple.
The non-trivial part of the volume operator is
\be
    \Vtwosq = \,\mid \Etwo \mid.
\label{V2}
\ee
In what
follows I will refer to \Vtwosq (or its square root) as the "volume"
operator, for short, even though \Vtwo has the dimensions of an
area. (I cannot call \Vtwo, e.~g. transverse area; \E{z}{Z} is the
transverse area.)

    The wavefunction is actually a product of two coherent
states, one for the x direction and one for the y direction.
\[
    \ket{u,\vec{p}}\rightarrow \:\ket{u_x,\vec{p}_x;u_y,\vec{p}_y}
                = \:\ket{u_x,\vec{p}_x}\ket{u_y,\vec{p}_y}.
\]
The volume operator contains both \E{A}{x}, which acts only on the
first ket, and \E{B}{y}, which acts only on the second ket.
From the discussion of the action of $E^a$ on a coherent state,
paper 1, the SC contributions have the form
\begin{gather}
    (\kappa\gamma/2)^{-1}E^x_A \ket{u_x,\vec{p}_x} = N\{ <L> \hat{L}_A\sum_L\cdots(1) \nn
            + (i \hat{L} \times \hat{D} + \hat{D})_A \sum_L \cdots (M)+ \hat{L}_A \,\sum_L \cdots(L-<L>)\} \nn
            = \, <L> \hat{L}_A \,\ket{u_x,\vec{p}_x} + [\,N/N(1M)\,] \,(i \,\hat{L} \times \hat{D} + \hat{D})_A \ket{1M}\nn
            + [\,N/N(1L)\,] \,\hat{L}_A \, \ket{1L}.
\label{VectorsxMoments}
\end{gather}
The $\cdots$ indicates parts of each summand  that
are the same for each sum.  $\hat{L}$ and $\hat{D}$
are the images of $\hat{p}$ and $\hat{Z}$, respectively,
under rotation by the peak holonomy u.  Since $\hat{p}$ and $\hat{Z}$
are perpendicular, their images $\hat{L}$ and $\hat{D}$
are perpendicular.  $\hat{L} \times \hat{D}$ completes the trio
of orthonormal vectors.  ($\hat{L}$ is also the direction of
peak L, \eq{Eeigenval}.)

    The N's are normalization factors.
\begin{align}
     1 &\cong  N^2 \exp[ \,p^2/t -p \,] \,\sqrt{(<L+1/2>/t)}/2\pi ;\nn
     N/N(1M) &\cong \sqrt{<L+1/2>/2}; \nn
     N/N(1L) &\cong \sqrt{<L+1/2>/2\,p)}.
            \label{NoverNi}
\end{align}

    Since the leading term in \eq{VectorsxMoments} is multiplied by $<L>$, the
moment states are down by factors of $1/\sqrt{<L>}$.
This suppression comes about because the normalization constant
N(1X) of a first moment state involves the
standard deviation.
 \[
    N(1X)^{-2} = \int X^2 \exp [- X^2/\sigma(X)^2] \sim \sigma(X)^2.
 \]
\[
    N(1X) \sim 1/\sigma \sim 1/\sqrt{L}.
\]

    The action of \Etwo on the coherent states has a form given
by \eqs{natlbasis1}{natlbasis2}, generalized to a
product of x and y coherent states.

 \begin{multline}
    (\gamma\kappa/2)^{-2} \Etwo \ket{u_x,\vec{p}_x}\ket{u_y,\vec{p}_y} =
        A(0)\ket{u_x,\vec{p}_x}\ket{u_y,\vec{p}_y}\\
        + \sum_{Xy} A(1Xy)\ket{u_x,\vec{p}_x}\ket{1X_y}\\
        + \sum_{Xx} A(1Xx)\ket{1X_x}\ket{u_y,\vec{p}_y} \\
        + \sum_{XxXy} A(1Xx,1Xy)\ket{1X_x}\ket{1X_y}.
            \label{defAmn}
 \end{multline}
 As at \eq{natlbasis1}, each \Etld produces only the original coherent
 state plus first moment SC states.  From
 \eq{defSC}, the leading term is order $<L_x L_y>$,
 \bea
       A(0) &=&  <L_x> <L_y> \:\hat{Z}\cdot \hat{L}^{(x)} \times \hat{L}^{(y)}\nn
            &\sim & <L_x> <L_y> \times \, \mbox {order unity}.
                    \label{A0}
 \eea
\begin{eqnarray}
    A(1Lx) &=&  \sqrt{<L_x>/2 \,p}\:\hat{Z}\cdot \hat{L}^{(x)} \times  \hat{L}^{(y)} <L_y>\nn
            &=& A_{1Ly}; \nn
    A(1Mx)(x,y) &=& \sqrt{(<L_x> )/2}\:\hat{Z}\cdot [\hat{D} + i \hat{L} \times \hat{D}]^{(x)}]\times \hat{L_y}^{(y)} <L_y>;\nn
    A(1My)(x,y) &=& - A(1Mx)(y,x);\nn
\label{defAmnX}
\end{eqnarray}
Since we are working to lowest order in $1/\sqrt{<L>}$,
we are done; the remaining coefficients are higher order.
 \begin{eqnarray}
    A(1Xa) &=& \mbox{order} A(0) (1/\sqrt{<L_a>}) \nn
    &=& \mbox{order}\sqrt{<L_a>}<L_b>;\nn
    A(1Xx,1Xy) &=& \mbox{order} A(0)(1/\sqrt{<L_x L_y>}) \nn
    &=& \mbox{order}\sqrt{<L_x L_y>}
        \label{omAmn}
\end{eqnarray}
a,b = x,y; a $\neq$ b.

    The action of \Etwo  on other members of the complete set follows
from \eq{natlbasis2}, for example
\begin{multline}
 (\gamma\kappa/2)^{-2}\Etwo \ket{1M_x}\ket{u_y,\vec{p}_y} =
         A(0)\ket{1M_x}\ket{u_y,\vec{p}_y} \\
  + \sum_{Xy}\tilde{b}(1Xy)\ket{1M_x}\ket{1X_y} \\
          + \tilde{b}(1M)\ket{2M_x}[ \,\ket{u_y,\vec{p}_y}+ \\
          + \tilde{b}(1MX)\ket{1M_x, 1X_x} \ket{u_y,\vec{p}_y}\\
          + c(1) \ket{u_x,\vec{p}_x}\ket{u_y,\vec{p}_y} \\
          + \cdots.
         \label{defBmn}
 \end{multline}
The $\cdots$ denotes states identical to the
preceding three lines, except y ket
\[
    \ket{u_y,\vec{p}_y} \rta  \ket{1X_y}.
\]
I will not need the $\tilde{b}$, if working only
to first order.    However, \eq{defBmn} implies that the
moment states are also approximate
eigenfunctions of \Etwo  with the
same eigenvalue (same leading coefficient $ A_0$).
This coefficient is not suppressed and will play a role
later.

\section{The Operator \Vtwo}
\label{V2}

        The last section constructed \Etwo; we must now
multiply this by a phase $\pm 1$ to obtain a magnitude, \Vtwosq,
then take a square root.   Although the original coherent state
and the moment states are not exact eigenvectors of \Etwo,
they are approximate eigenvectors.  They share the same
approximate eigenvalue, A(0).   $\mid \Etwo \mid$
has the same eigenvectors as \Etwo, provided we
maintain the phase relationships between the leading
term A(0) and the subsidiary terms  in \eqs{defAmn}{defBmn}. I.~e.,
only the overall phase can be changed in \eqs{defAmn}{defBmn}.

    I choose
the overall phase to be sgn [A(0)], in order to make the
A(0) term positive.  All states in the complete set are now
approximate eigenvectors of \Vtwosq, with eigenvalue A(0) sgn [A(0)]

    I assume that \Vtwo has the same non-zero
matrix elements as \Vtwosq.  \Vtwosq connects an initial
state, moments ($pX_x, qX_y$), to final states with
moments (p or p$\pm 1$, q or  q$\pm 1$).  I assume \Vtwo does
the same.   I assume \Vtwo, like \Vtwosq, is
a double power series in $1/\sqrt{L_a}$, a = x,y.
(If I assume, e.~g. \Vtwo = double power series in
$\sqrt[4]{<L_a>}$, then \Vtwosq would not be a power
series in $\sqrt{<L_a>}$.)
The matrix elements for \Etwo, \eqs{defAmn}{defBmn} have  diagonal
elements A(0) with $L_x L_y$ as
highest power; therefore the \Vtwo  series
has diagonal elements a(0) with highest power $\sqrt{L_x L_y}$.

   The ansatz for \Vtwo  is then
\begin{gather}
    (\gamma\kappa/2)^{-1}\Vtwo \ket{u_x,\vec{p}_x}\ket{u_y,\vec{p}_y}=
        a(0)\ket{u_x,\vec{p}_x}\ket{u_y,\vec{p}_y} \nn
         + \sum_{Xy} d(1Xy)\ket{u_x,\vec{p}_x}\ket{1X_y}\nn
        + \sum_{Xx} d(1Xx)\ket{1X_x}\ket{u_y,\vec{p}_y} ;\nn
    (\gamma\kappa/2)^{-1}\Vtwo \ket{1X_x}\ket{u_y,\vec{p}_y} =
        a(0)\ket{1X_x}\ket{u_y,\vec{p}_y} \nn
        + \tilde{d}(1Xy)\ket{1X_x}\ket{1X_y} \nn
        + \tilde{d}(1X'x)\ket{1X_x, 1X'x \mbox{or} 2X}\ket{u_y,\vec{p}_y} \nn
        + c(1)\ket{u_x,\vec{p}_x}\ket{u_y,\vec{p}_y} \nn
        + \cdots
\label{V2Ansatz}
\end{gather}
and similarly for the action of \Vtwo on
$\ket{u_x,\vec{p}_x}\ket{1(X_y}$.
\begin{gather}
    a(0) = \mbox{order} \sqrt{L_x L_y};\nn
     d(1Xa), \: \tilde{d}(1Xa), \: c(1) = \mbox{order}\sqrt{L_x L_y} (\sqrt{1/L_a})^{-1}.
        \label{OMab}
\end{gather}
The terms indicated by $\cdots$ have y kets
\[
    \ket{u_y,\vec{p}_y} \rta \ket{1X_y},
\]
as at \eq{defBmn}.

   To determine the $a(0)$, I set

\be
    \Vtwosq \ket{u_x,\vec{p}_x}\ket{u_y,\vec{p}_y}=
            \sum_n \Vtwo \ket{n} \bra{n}\Vtwo \ket{u_x,\vec{p}_x}\ket{u_y,\vec{p}_y}.
                \label{Aeqasq}
\ee
The left-hand side is an expansion in A's, which are known; the
right-hand side will be an expansion quadratic in a's, d's, and c,
the unknowns.

    To determine a(0), I multiply \eq{Aeqasq} from the left by
the original coherent state
\[
    \bra{u_x,\vec{p}_x}\bra{u_y,\vec{p}_y}.
\]
This projects out $A(0)$ on the left. On the right, if we are keeping
only zeroth plus first order, the only intermediate state n which
contributes is the original coherent
state.  Intermediate states $\ket{n}$ with one first moment
ket do not contribute, because a product
of two d(1Xa) matrix elements, \eq{OMab}, is down by order
$1/<L_a>$.  Therefore
\begin{eqnarray}
    A(0) \,sgn \,[ \,A(0) \,] &=& a(0)^2 \,(1  + \mbox{order} \,1/<L>); \nn
    a(0) &=& \sqrt{A(0) \,sgn \,[ \,A(0) \,] \,}(1 + \mbox{order} \,1/<L>).
\label{a0}
\end{eqnarray}
The first order corrections to $a(0)$ vanish.

    Once $a(0)$ is determined, the remaining d(1Xa) in \eq{V2Ansatz}
follow.  For example, multiplying \eq{Aeqasq} from the left by
the first moment ket $\bra{u_x, \vec{p}_x, 1Xy}$ projects
out A(1Xy) on the left.
\[
    A(1Xy) = \bra{u_x, \vec{p}_x, 1Xy}Vtwo\ket{n}\bra{n}Vtwo\ket{u_x,\vec{p}_x}\ket{u_y,\vec{p}_y}.
\]
The right-hand side can be order $\sqrt{1/<L>}$  only if
one of the two matrix elements is not suppressed, i.~e. one
of the two matrix elements must be $a(0)$.  From
\eq{V2Ansatz}, that matrix element must be on-diagonal.  The intermediate
state n must  be either the original
coherent state, or a 1Xy state.
\begin{eqnarray}
    A(1Xa) \,sgn \,[ \,A(0) \,] &=&  [ \,d(1Xa)\, a(0) + a(0)\, d(1Xa) \,] \nn
        && \times (1+ \mbox{order}\sqrt{1/<L>}) ;\nn
    d(1Xa) &=& \{A(1Xa) \,sgn\,[ \,A(0) \,]/( \,2\sqrt{A(0)\, sgn \,[ \,A(0) \,] \,} \,)\} \nn
            && \times (1 + \mbox{order} \sqrt{1/<L>}).
\label{d1}
\end{eqnarray}
The first order corrections are now determined.

    The process also determines higher order corrections.  Consider elements
\[
    \bra{i}\Vtwosq\ket{i} = \bra{i}\Vtwo\ket{n}\bra{n}\Vtwo\ket{i};
\]
\[
    \bra{f}\Vtwosq\ket{i} = \bra{f}\Vtwo\ket{n}\bra{n}\Vtwo\ket{i}.
\]
On the first line, only diagonal elements n = i contribute to
lowest order.  As at \eq{a0}, one obtains a(0) = $\sqrt{A(0) \,sgn\,[A(0)]}$.
On the second line, f $\neq$ i can be i plus or minus one or
two additional moments.  For example, in
\begin{multline}
    \bra{pXx, (q+1)Xy}\Vtwosq\ket{pXx,qXy} \\
    = \bra{pXx, (q+1)Xy}\Vtwo\ket{n}\bra{n}\Vtwo\ket{pXx,qXy}
\label{fVVi}
\end{multline}
the highest contributions come from
\begin{eqnarray*}
    \ket{n} &=& \ket{f} \,\mbox{or} \ket{i} \\
            &=& \ket{pXx, (q+1)Xy} \,\mbox{or} \ket{pXx,qXy}.
\end{eqnarray*}
\Eq{fVVi} becomes
\begin{multline}
    \bra{pXx, (q+1)Xy}\Vtwosq\ket{pXx,qXy} \\
    = 2\bra{pXx, (q+1)Xy}\Vtwo\ket{pXx,qXy} \,a(0) .
\end{multline}
In this manner unknown, off-diagonal elements of \Vtwo are expressed
in terms of known, diagonal elements.

\section{The Inverse of \Vtwo}
\label{V2Inverse}

    The matrix \Vtwo  has a kernel, and cannot
have an exact
inverse.  However, if \Vtwo  acts on a coherent state which is an
approximate eigenvector of \Vtwo with very large eigenvalue, then
one can neglect the states in the kernel, because \Vtwo  connects the
original state only to other (original and moment) states
having large eigenvalue: note the structure of \eq{V2Ansatz}.

    In this situation it is possible to calculate an approximate
inverse for \Vtwo.  As in the calculation of \Vtwo, I
work only to zero and first order in $1/\sqrt{\mathrm{L}}$.

    The matrix elements of \Vtwo  were assumed to be
order $\sqrt{L_x L_y}$, times a power series in powers of
$1/\sqrt{L_a}$.  This suggests a corresponding ansatz for
matrix elements of
$(\Vtwo)^{-1}$: they are order $1/\sqrt{L_x L_y}$, times a
power series in powers of $1/\sqrt{L_a}$.

    Also, the approximate inverse must obey
\begin{multline}
    \delta(ix.jx) \,\delta(iy,jy) = \Sigma_{nx,ny}
    \bra{jx,jy}(\Vtwo)^{-1}\ket{nx,ny} \\
    \cdot \,\bra{nx,ny}\Vtwo\ket{ix,iy}.
        \label{defVinv}
\end{multline}
The indices (ix, iy), (jx, jy), (nx, ny) label states in
the complete set, \eq{completeset}.

    As a first step I keep only the leading, zeroth order,
contributions to
the matrix \Vtwo.  In zeroth order \Vtwo  is diagonal:
\[
    \bra{nx,ny}\Vtwo\ket{ix,iy} = \delta (nx,ix) \,\delta(ny,iy) \,a(0).
\]
Therefore the matrix $(\Vtwo)^{-1}$ is also diagonal.  To zeroth
order,

\be
    \bra{jx,jy}(\Vtwo)^{-1}\ket{nx,ny} = \delta (jx,nx) \,\delta (jy,ny)/a(0).
        \label{Vinv0}
\ee

  Next, I include first order contributions to \Vtwo.  In
the sum over (nx,ny), \eq{defVinv}, I can drop all products of matrix
elements which are second order or higher.  Then the matrix elements in
\eq{defVinv} must be diagonal, with each ket index equal to the
corresponding bra index (ix = nx, nx = jx, etc.), except for one
pair, which I will call the off-diagonal
pair. For example, suppose the
off-diagonal pair is ix $\neq$ jx; then the requirement of at most
one off-diagonal matrix element implies only two values of nx are
possible; either ix = nx but jx $\neq$ nx; or the reverse, ix $\neq$ nx
but jx = nx. Any other choice of nx would lead to a product of two
first order matrix elements, therefore a contribution to the sum
\eq{defVinv} which is second order.  Also, if the
off diagonal pair is an x-pair (as in our example, where ix $\neq$ jx)
then all y dependence must be diagonal to avoid second order contributions:
iy = ny = jy.

    For a specific example with ix $\neq$ jx, consider
ix = the 1M state, jx = the original coherent state.  \Eq{defVinv}
becomes

\bea
    0 &=&  \Sigma_{nx}
    \bra{u_x,\vec{p}_x;\cdots}\Vtwo^{-1}\ket{nx,\cdots}\bra{nx,\cdots}\Vtwo\ket{1M_x;\cdots}\nn
    &=& \bra{u_x,\vec{p}_x}\Vtwo^{-1}\ket{u_x,\vec{p}_x}\bra{u_x,\vec{p}_x}\Vtwo\ket{1M_x}\nn
     & &   +
        \bra{u_x,\vec{p}_x}\Vtwo^{-1}\ket{1M_x}\bra{1M_x}\Vtwo\ket{1M_x}.
        \label{Vinvcalc1}
\eea
The $\cdots$ on the first line indicate suppressed labels iy, ny, jy
which are all equal and
do not change. The intermediate state $\ket{nx}$ = \mbox{$\ket{1L_x}$}
does not contribute; matrix elements
\[
    \bra{1L} \Vtwo \ket{1M}
\]
are second order.  (Note \emph{two} changes in moment: 0L $\rta$ 1L;
1M $\rta$ 0M.)

    Since three out of four of the matrix elements in \eq{Vinvcalc1}
are known to the required order, I can solve for the fourth matrix
element.

\begin{multline}
    \bra{u_x,\vec{p}_x;iy} \Vtwo^{-1} \ket{1M_x;iy} \\
    = -\bra{1M_x; iy} \Vtwo \ket{u_x,\vec{p}_x; iy}^*/a(0)^2 \\
    = -d^*(1M_x)/a(0)^2.
    \label{Vinvcalc2}
\end{multline}
By systematically considering all $ix\neq jx$ and $iy \neq jy$
pairs, I can determine all first order elements of the inverse.

    \Eq{Vinv0} for the diagonal elements of
$(\Vtwo)^{-1}$ is already correct to first order; there are no
first order corrections to the diagonal elements.  Proof: the diagonal
elements have ix = jx and iy = jy.  If also ix = nx and iy = ny,
then all matrix elements are diagonal and we need a first order
correction to a diagonal element.  There are none (see \eq{Aeqasq}).
If (say) ix $\neq$ nx, then jx $\neq$ nx also.
This contribution to the sum over $n_x$ is second order, and may be
dropped from the sum.  \Eq{Vinv0} for the diagonal elements of
$(\Vtwo)^{-1}$ is already correct to first order. $\Box$

    To first order, the $(\Vtwo)^{-1}$ matrix is 5 x 5, connecting the five states
\begin{gather}
    \ket{u_x,\vec{p}_x}\ket{u_y,\vec{p}_y},\ket{u_x,\vec{p}_x}\ket{1Ly},\ket{u_x,\vec{p}_x}\ket{1My} \nn
        \ket{1Lx}\ket{u_y,\vec{p}_y}, \ket{1Mx}\ket{u_y,\vec{p}_y}.
\end{gather}
The matrix has the form
$$(\Vtwo)^{-1} = (1/a(0))\begin{bmatrix}
            1 & -U \\
            -U^{\dag} & 1_4
            \end{bmatrix}
$$
U is the four dimensional row vector
\[
    U = [ \,d(1Ly), \,d(1My), \,d(1Lx), \,d(1Mx) \,]^*/a(0).
\]
$1_4$ is a 4 x 4 unit matrix.
The eigenvalues are 1/a(0) (degeneracy three) and
\[
    (1 \pm \sqrt{ \,\sum_{a,X}|\,d(1Xa)\,|^2/a(0)^2\,})/a(0).
\]
The three degenerate eigenvectors have the form
\[
    [\,0,\,W\,].
\]
W is a four dimensional row vector obeying $W^{\dag} \cdot U = 0$.
The remaining two eigenvectors have the form
\[
    [\,\mp \sqrt{ \,\sum_{a,X}|\,d(1Xa)\,^2\,}/a(0), \,U^* \,].
\]

\section{Conclusion}\label{SecConclusion}

    This paper constructed a complete subset of the overcomplete
set of coherent states, and used it to construct a
perturbation expansion for the inverse of \Vtwo.
The subset consists of the original coherent
state, which has Gaussian distributions in L and M, plus
SC states with
distributions given by moments of the Gaussians.  All
the formulas giving the action of the holonomy and \Etld
operators on the coherent
state may be considered as perturbation expansions in this
complete subset.

    The technique works quite generally.
One can use moments to construct
a complete subset,  for calculations involving any kind of coherent
state.

    Is this perturbation approach useful for extending coherent
state calculations to smaller values of L?  Here we are limited
by the slow falloff of higher
terms in the perturbation series, as $1/\sqrt{<\mathrm{L}>}$
rather than 1/L.  In the neighborhood of $<\mathrm{L}> \sim$ 100 the
higher terms are only a 10\% correction, but by $<\mathrm{L}> \sim$ 10
the series is converging very slowly.

    For small L there is also a possible problem with spreading of the
coherent wave packet.  Coherent states are
useful because their eigenvalues are strongly peaked at one central
value.  If this peak broadens rapidly in time, the states lose their
coherent character.  In appendices \ref{WeakSpreading} and D I estimate the
rate of spreading of a coherent state, and for small L the rate of
spreading increases as 1/$\sqrt{<\mathrm{L}>}$.

\appendix

\section{Fluctuations in Angles}\label{AppAngles}

    Given that the matrix h is strongly peaked at the value u
it should be possible to translate this into a statement
that the angles ($\theta, \phi$) are strongly peaked at
($\alpha, \beta$), then obtain an estimate for the standard
deviations of these angles.

    To begin, establish the two expectation values

\begin{gather}
    0 = \bra{u,\vec{p}}(D^{(1)}(h)- D^{(1)}(u))_{0A}\ket{u,\vec{p}};
        \label{meanh}\\
    \mbox{order}\, (p/<L>),1/<L> = \bra{u,\vec{p}}(D^{(1)}(h)\dag -
    D^{(1)}(u)\dag)_{B0} \nn
          \cdot (D^{(1)}(h)-D^{(1)}(u))_{0A}\ket{u,\vec{p}}.
            \label{stndevh}
\end{gather}
The first equation is related to a
mean; the second to a standard deviation.  The first equation is
zero because $(D(h) - D(u))\ket{u,\vec{p}}$ contains only SC
states, and the leading SC states are odd in one of the
variables, M or $L
-<L>$, while $\bra{u,\vec{p}}$ is even.

    What happens when a second factor of D(h) - D(u) is applied
to the state, as in \eq{stndevh}?  The right hand factor
$(D^{(1)}(h)-D^{(1)}(u))_{0A}\ket{u,\vec{p}}$ is the sum of SC states
given in paper 1, section on action of the holonomy

\bea
    \ket{SC_A}&:=&(D^{(1)}(h)-D^{(1)}(u))_{0A}\ket{u,\vec{p}}\nn
           & =& [\,N/N(L+1,1M) \,]\ket{L+1,1M} \nn
           & & \cdot [\, -\hat{L}_A/(2L+1) \,] \nn
           & & -i [\,N/N(L_-) \,]\ket{L_-} [ \,\hat{D} \times \hat{L} \,]_A.
           \label{defSCA}
\eea
D(h) acting on a state with peak angular momentum L produces
states with peak angular momentum L $\pm$ 1, with or without
mmoments X inserted in the summand.  N($L_-$) is the normalization
constant for
\[
    \ket{L_-} = [\,\ket{L+1} - \ket{L-1} \,]/2
\]
which is not a moment state but nevertheless is very small because
it is the difference of two very similar Gaussians.
\[
    \bra{u,\vec{p}}(D^{(1)}(h)\dag- D^{(1)}(u)\dag)_{B0}
\]
is the Hermitean conjugate $\bra{SC_B}$.  The dot product of the two
factors gives vector components of order unity times factors of order

\[
    \{N/[N(L+1,1M)(2L+1)]\}^2, \, [N/N(L_-)]^2
\]
which are order 1/L and p/L, respectively, from paper 1.$\Box$

    In \eq{stndevh}, the left-hand side is now known.

\begin{gather}
    \braket{SC_B}{SC_A} = \bra{u,\vec{p}}(D^{(1)}(h)\dag-
                    D^{(1)}(u)\dag_{B0}) \nn
            \cdot (D^{(1)}(h)-D^{(1)}(u))_{0A}\ket{u,\vec{p}},
            \label{stndevh2}
\end{gather}
On the right-hand side, I expand D(h) around D(u).  Define
\[
    \delta \theta = \theta - \alpha; \,
    \delta \phi = \phi - \beta.
\]
Then

\bea
    D(h)_{0A}&=& d(\alpha/2+\delta \theta/2)_{0A}\exp[i S_Z (\beta +
    \delta \phi  -\pi/2)]\nn
            &=&d(\alpha/2)_{0N}\exp[i S_Z (\beta -\pi/2)]\exp[i(\delta\theta/2) \vec{S}\cdot
            \hat{n}]_{NA}\exp[i S_Z (\delta \phi )]\nn
            &\cong&D(u)_{ON}[1 +i (\delta\theta/2) \vec{S}\cdot
            \hat{n}+ i S_Z (\delta \phi)]_{NA}.
            \label{Dhexpan}
\eea
The second line uses the behavior of the y axis under rotation
around Z,

\[
    \exp[-i S_Z (\beta-\pi/2] \,S_Y \exp[i S_Z(\beta-\pi/2)] = \vec{S}\cdot\hat{n}.
\]
Now insert \eq{Dhexpan} into \eq{stndevh2}:

\begin{gather}
    \braket{SC_B}{SC_A} \cong \bra{u,\vec{p}}\{D(u)[ (\delta\theta/2) \vec{S}\cdot
            \hat{n}+  S_Z (\delta \phi)]\}_{0A} \nn
             \cdot \{[ (\delta\theta/2) \vec{S}\cdot
            \hat{n}+  S_Z (\delta \phi)]D(u)\dag\}_{B0}\ket{u,\vec{p}}.
            \label{stnddev3}
\end{gather}
All matrices D(u) have L = 1.  On the right, the $(\delta \theta \, \delta \phi)$ cross
terms average to zero, leaving a positive definite expression
linear in \mbox{$(<\delta
\theta)^2>$}, \mbox{$<(\delta \phi)^2>$}.

    To separate the two standard deviations, first take A = B
and sum over A.  The matrix D(u) has its axis of rotation along $\hat{n}$,
therefore
\[
    D(u)\, \vec{S}\cdot \hat{n}\, D(u)\dag = \vec{S}\cdot \hat{n}.
\]
D(u) rotates the Z axis into the vector $\hat{D}$,
therefore
\[
    D(u)\, S_Z \,D(u)\dag = \vec{S}\cdot \hat{D}.
\]
The l.h.s. of \eq{stnddev3} comes from the norm of \eq{defSCA}.
\bea
    \mbox{l.h.s.}&=&\{N/[N(L+1,m1(M))(2L+1)]\}^2 + [N/N(L_-)]^2;\nn
    \mbox{r.h.s}&=& [<(\delta\theta/2)^2>+\sin^2(\alpha/2)<(\delta \phi)^2>].
    \label{stnddev4}
\eea
The right-hand side uses (remember the $\vec{S}$ are 3x3)
\begin{eqnarray*}
    [(\vec{S}\cdot\hat{D})^2]_{00} &=& [L(L+1)/2](D_X^2 + D_Y^2)\\
            &=& [L(L+1)/2]\sin^2(\alpha/2),
\end{eqnarray*}
plus L = 1; and a similar formula for $(\vec{S}\cdot\hat{n})^2$, except
$n_X^2 + n_Y^2= 1$.

    Next take A = B
= 0  in \eq{stnddev3}.  This kills the $\delta \phi$ terms.  In \eq{defSCA}
the $\hat{n}$ terms disappear, since this vector has no Z component.  The
new left- and right-hand sides are

\bea
    \mbox{l.h.s.}&=&\sin^2(\alpha/2)\{N/[N(L+1,1M)(2L+1)]\}^2\sin^2 \mu \nn
                & &+ \cos^2 \mu \,[N/N(L_-)]^2;\nn
    \mbox{r.h.s}&=& [<(\delta\theta/2)^2>\sin^2(\alpha/2)].
    \label{stnddev5}
\eea
From paper 1 the normalizations are
\bea
    N/N(L_-) &\cong& \sqrt{p/2 \,<L+1/2>};\nn
    N/N(L+1,1M) &\cong& \sqrt{<L+1/2>/2} .
        \label{NoverNiforh}
\eea

    \Eqs{stnddev4}{stnddev5} may be solved for the individual
standard deviations.
\bea
    <(\delta\theta/2)^2> &=& \sin^2 \mu (1/(8<L+1/2>) \nn
                            & & + \cos^2\mu \,(p/2<L+1/2>)];\nn
    \sin^2(\alpha/2)<(\delta \phi)^2> &=& \cos^2 \mu (1/(8<L+1/2>) \nn
                                & & + \sin^2\mu \,(p/2<L+1/2>).
    \label{stnddev6}
\eea

    The $p/2<L+1/2>$ terms in \eq{stnddev6} are caused by
fluctuations in L.  To understand the structure of these
terms, construct a trio of orthogonal vectors: $\hat{D}$,
$\hat{n}$, and $\hat{n}\times \hat{D}$.   The angular momentum
$<\vec{L}>$ (peak value of angular
momentum) is perpendicular to $\hat{D}$:  $\hat{D}$ and
$<\vec{L}>$ are images of the vectors $\hat{Z}$ and $\vec{p}$, respectively, under the
rotation u, and $\hat{Z}$ and $\vec{p}$ are perpendicular; therefore
$\hat{D}$ and $<\vec{L}>$ are perpendicular.  The formula for $\hat{L}$,
\[
    \hat{L} = \hat{n} \cos \mu - \hat{n} \times \hat{D} \sin \mu,
\]
also confirms that $\hat{L}$ is perpendicular to $\hat{D}$.

     The single polarization constraints imply specific values for
$\mu$.  However, for the moment do not impose this constraint;
suppose $\hat{L}$ can be any combination of $\hat{n}$
and  $\hat{n} \times \hat{D}$ .  For $\hat{L}$ along $\hat{n}$
($\cos \mu$ = 1),
$\hat{D}$ must oscillate along $\hat{n} \times \hat{D}$;
for $\hat{L}$ along $\hat{n} \times \hat{D}$ ($\sin \mu$ = 1),
$\hat{D}$ must oscillate along $\hat{n}$.  These oscillation directions
preserve the orthogonality of $\hat{D}$ and $\hat{L}$.

    The $\mu$ dependence of \eq{stnddev6} is consistent with the
above picture.  For $\hat{L}$ along $\hat{n}$, $\cos \mu$ is unity,
and the $p/2<L+1/2>$ terms in \eq{stnddev6} produce only oscillations of
$\theta$, i.~e. oscillations along $\hat{n} \times \hat{D}$.
For $\hat{D}$ along $\hat{n} \times \hat{D}$, $\sin \mu$ is unity  ,
and the $(p/2<L+1/2>$ terms in \eq{stnddev6} produce only oscillations of
$\phi$, i.~e. oscillations along $\hat{n}$.

\section{SC States and Standard Deviations}\label{AppStndDev}

    The uncertainty principle does not allow the SC terms
to vanish. Consider, for example, the SC terms generated by the \Etld
operator.  If they vanish, then the coherent state is an exact
eigenstate of the spin.  (The \Etld operator brings down a factor
of spin, therefore is essentially the spin operator.) If spin is
exact, then from the uncertainty principle the canonically
conjugate variables, the angles $\theta,\phi)$, must be completely
uncertain.  This implies a completely uncertain holonomy, whereas
a  coherent state must have both spin and holonomy peaked.
Therefore the SC terms cannot vanish.

    One can make a more quantitative statement about the SC terms and
the uncertainty principle: the SC terms yield the standard deviations of
the operators \Etld and $h^{(1/2)}$ around their means.  Proof: let O
be an operator which is an approximate eigenvalue of the coherent
state,

\bea
    O_A \ket{u,\vec{p}} &=& <O_A> \ket{u,\vec{p}} + SC \nn
                    &=& <O_A> \ket{u,\vec{p}} + \sum_i \hat{e_i}_A
                    \lambda_i \ket{nc_i}.
                    \label{deflambda}
\eea
I have given O a vector subscript because the important operators
in this paper are vectors; but this feature is unimportant. The
$\hat{e_i}$ are unit vectors.  When the standard deviation is
computed, the original coherent state drops out, and only the near
coherent, SC terms contribute:

\bea
    \bra{u,\vec{p}}(O_A-<O_A>)^2\ket{u,\vec{p}} &=& \sum_{ij}\hat{e_j}_A^* \hat{e_i}_A
                    \lambda_j^* \lambda_i \braket{nc_j}{nc_i}\nn
                    &=&\sum_i \hat{e_i}_A^* \hat{e_i}_A
                    \lambda_i^* \lambda_i,
            \label{sdcalc1}
\eea
On the last line I have used the orthogonality of the near
coherent states.  From \eq{sdcalc1}, only the near coherent states
contribute to the total standard deviation squared.

     From the discussion of SC states in paper 1, each
$\lambda_i$ will contain a ratio of normalizations,
\[
    N/N(i) = \mbox{norm of original coh state/norm of SC state}.
\]
The above discussion clarifies why each
N/N(i) contains a factor which may be interpreted as a
standard deviation, for either the L or M distribution.
\Eq{sdcalc1} expresses the total squared standard deviation as
the sum of individual squared standard deviations
$\mid\lambda_i\mid^2 $ contributed by each near coherent state.
In statistics, the sum-of-squares form is obtained when the
fluctuations in a total quantity are caused by fluctuations
in several variables which are statistically independent.

\section{Packet Spreading: Weak Fields}
\label{WeakSpreading}

    In the weak field limit the gravitational Hamiltonian decouples
into a sum of oscillators.  These are especially easy to treat using
coherent state methods.  We can expect something like
\be
    <\delta \Etld(z,T=0)> = A\cos \,(k z +\zeta)
    \label{defzeta}
\ee
for time T = 0; and oscillator packets are known to follow the classical path
exactly for T$>$0 \cite{Glauber}:
\[
    <\delta \Etld(z,T)> = A\cos \,(k z -\omega T +\zeta).
\]
$\delta \Etld$ is the fluctuation of \Etld away from flat space.

    Although the \emph{average} value of oscillator displacement follows
the classical path, this is not enough to guarantee classical behavior.
The fluctuations around the average must also be small.  For the
usual oscillator these fluctuations are determined by $\delta \zeta$,
the uncertainty in the phase $ \zeta$ introduced at \eq{defzeta}.
This uncertainty is connected to the uncertainty in the number of quanta by
$\delta N \,\delta\zeta \sim 1$, which leads to
 $\delta\zeta \sim 1/\sqrt{N}$, since N has
standard deviation $\sqrt{N}$.

    In the LQG case presumably $\zeta$ will be a function of the
dimensionless angles $(\alpha,\beta)$ used to define the
unitary matrix u, as well as
the angles defining the unit vector $\hat{p}$. ($\zeta$
could also be a function of a dimensionless ratio of angular momenta,
(average Z component)/ $<L>$; but this ratio depends on
the angles already listed.)
The standard deviations for fluctuations in the angles are order
$1/\sqrt{<L>}$.  $<L>$ therefore replaces N; and for L of order
100 or so, the  fluctuations around the classical path should be
small.

\subsection{Spreading in the Strong Field Limit}
\label{StrongSpreading}

    Once the self-interaction of the gravitational packet is
included, it becomes much harder to estimate the rate of spreading.
Consider two familiar quantum mechanical examples, the free
particle and the simple harmonic oscillator.   The spreading of the free
particle wave packet is governed  by the time scale

\be
    T_f = m (\sigma_f)^2/\hbar,
        \label{FreeSpread}
\ee
where m is the mass and $\sigma_f$ is the standard
deviation of the T = 0 Gaussian packet in configuration space.

    At the other extreme, the Gaussian packet for the simple harmonic oscillator
does not spread at all \cite{Glauber}.  Evidently the rate of
spreading is highly sensitive
to the details of the energy spectrum \cite{Klauder}.

    The planar case, as well as the general SU(2) case, has
an asymptotic region, therefore a
Hamiltonian \cite{Regge}; and it makes sense to talk about energies E.
If the
energy eigenvalues are evenly spaced, like those of the
oscillator, then the likelihood of spreading should be small.

    The Hamiltonian is a surface term.  Its size can be estimated,
using dimensional analysis. The surface term can be rewritten as a
density.  In the planar case,

\[
    E|^{+\infty}_{-\infty} = \int \partial_z E \,dz.
\]
E is also integrated over a transverse area in the
xy plane (not shown), so that the integral on the right
is a volume integral. Since the transverse area is not
physically meaningful, I take it to be a unit area.

    I assume the volume integral on the right resembles the
typical terms in the Euclidean Hamiltonian.
\be
    E \sim \epsilon^{ijk} \,Tr(h_{ij} \,h_k \,[ \,h_k^{-1},V \,] \,)/\kappa^2 .
    \label{EestI}
\ee
(I could also use the
terms in the Hamiltonian which are proportional to the square of the
extrinsic curvature; the order of magnitude estimates would be the same.)
I need to estimate the dependence of this expression on angular momentum
L.  The volume V contains three \Etld operators, integrated over
area, with area eigenvalues of order $L \kappa$.  Therefore volume
V should be order $(L \kappa)^{3/2}$.  However, the commutator of
the volume with holonomy takes the derivative of V with respect to
L; see the discussion of the commutator in paper 1.
Therefore $h_k \,[h_k^{-1},V]$  is order
$\sqrt{L}(\kappa)^{3/2}$.
The remaining holonomy in the Hamiltonian give a result of order
unity when acting on a state.  Therefore the energy grows as the
square root of L.

\be
    E \sim \sqrt{(L \kappa)}/\kappa.
    \label{EestII}
\ee
This resembles the classical expression for the gravitational
Hamiltonian,
\[
     \mbox{(curvature)}\, d^3x /\kappa \simeq \mbox{(large
length)}/\kappa,
\]
except that the large length has been replaced
by the square root of an area eigenvalue.  As a further check, the
mass and event horizon area of a black hole are connected by
a relation of similar form,
mass $\propto$ square root of area.

    For minimal spreading, the spacing between energy levels should be
as constant as possible, resembling the spacing between levels of the
usual oscillator \cite{Klauder}.  From \eq{EestII} the spacing is order

\be
        \delta E \sim \hbar c \, \delta L /\sqrt{L \kappa} := \delta L \,\hbar \,\omega
        \label{EestIII}
\ee
I have restored factors of $\hbar$ and c, and given $\kappa$ the dimensions
of a length.

    $\delta L \,\hbar \,\omega$ resembles the SHO
formula $\delta n \,\hbar \,\omega$  Is $\hbar \,\omega$ a constant?
\be
        \hbar \,\omega =  \hbar \,c /\sqrt{L \kappa}.
               \label{defw}
\ee
At first glance this result is not what we want: the quantity $\omega$ is not a
constant, independent of L.  However, $\omega$ does not need to be a
constant everywhere; it must be an approximate constant for the range of L
values contained in a coherent state.  Over this range, the fractional
change in $ \omega$ is of order
\begin{equation}
        \delta \omega/\omega = - \delta L/(2 L) = - (1/\sqrt{t})/(2 L).
\label{dwoverwI}
\end{equation}
I estimate $\delta L$, the range of L values in the coherent state, using
the standard deviation of the Gaussian, $\sqrt{1/t}$.
As discussed at table \ref{Ta:pvsL}, the parameter $\sqrt{1/t}$ must be
small compared to L.  Alternatively, t is given exactly by  $<L>$ = p/t.  I insert
this value into \eq{dwoverwI}, replacing the L in that formula
by $<L>$, and get
\begin{equation}
    \delta \omega/\omega \sim - \sqrt{p/<L>}).
    \label{dwoverwIII}
\end{equation}
Even in the strong field case, a value $<L> \,\geq 100 p$ should be enough
to drive $\delta \omega$ to zero and prevent
spreading.

    \Eq{dwoverwI} is yet another reason why t cannot be made arbitrarily small:
the spacing between levels would no longer be uniform over the packet, leading
to unacceptable spreading.  It is perhaps relevant that $1/\sqrt{t}$ plays the
role of a standard deviation in spin network theory.  The
time scale $T_f$ for spreading of the free particle packet is also sensitive
to a standard deviation, $\sigma_f$; see \eq{FreeSpread}.

\end{document}